\documentclass[%
 aip,
floatfix,
amsmath,amssymb,
reprint,%
]{revtex4-1}

\usepackage{graphicx}
\usepackage{dcolumn}
\usepackage{bm}

\usepackage[utf8]{inputenc}
\usepackage[T1]{fontenc}
\usepackage{mathptmx}
\usepackage{etoolbox}
\usepackage[colorlinks, citecolor={blue}, urlcolor={blue}, linkcolor={blue}]{hyperref}

\newcolumntype{L}{>{\centering\arraybackslash}m{.15\textwidth}}

\makeatletter
\def\@email#1#2{%
 \endgroup
 \patchcmd{\titleblock@produce}
  {\frontmatter@RRAPformat}
  {\frontmatter@RRAPformat{\produce@RRAP{*#1\href{mailto:#2}{#2}}}\frontmatter@RRAPformat}
  {}{}
}%
\makeatother

\begin{document}

\preprint{AIP/123-QED}

\title{Ion Trap with In-Vacuum High Numerical Aperture
Imaging for a Dual-Species Modular Quantum Computer}
\author{Allison L. Carter}
\altaffiliation[Current address: ]{National Institute of Standards and Technology, Boulder, CO, 80305}
\affiliation{Joint Quantum Institute and Department of Physics, University of Maryland, College Park, 20742}
\email{allison.carter@nist.gov}

\author{Jameson O'Reilly}%
\affiliation{Joint Quantum Institute and Department of Physics, University of Maryland, College Park, 20742}
\affiliation{Duke Quantum Center, Department of Electrical and Computer Engineering, Department of Physics, Duke University, Durham, NC 27701}
\author{George Toh}
\affiliation{Joint Quantum Institute and Department of Physics, University of Maryland, College Park, 20742}
\affiliation{Duke Quantum Center, Department of Electrical and Computer Engineering, Department of Physics, Duke University, Durham, NC 27701}
\author{Sagnik Saha}
\affiliation{Joint Quantum Institute and Department of Physics, University of Maryland, College Park, 20742}
\affiliation{Duke Quantum Center, Department of Electrical and Computer Engineering, Department of Physics, Duke University, Durham, NC 27701}
\author{Mikhail Shalaev}
\affiliation{Duke Quantum Center, Department of Electrical and Computer Engineering, Department of Physics, Duke University, Durham, NC 27701}
\author{Isabella Goetting}
\affiliation{Duke Quantum Center, Department of Electrical and Computer Engineering, Department of Physics, Duke University, Durham, NC 27701}
\author{Christopher Monroe}
\affiliation{Joint Quantum Institute and Department of Physics, University of Maryland, College Park, 20742}
\affiliation{Duke Quantum Center, Department of Electrical and Computer Engineering, Department of Physics, Duke University, Durham, NC 27701}
\date{\today}

\begin{abstract}
Photonic interconnects between quantum systems will play a central role in both scalable quantum computing and quantum networking. Entanglement of remote qubits via photons has been demonstrated in many platforms; however, improving the rate of entanglement generation will be instrumental for integrating photonic links into modular quantum computers. We present an ion trap system that has the highest reported free-space photon collection efficiency for quantum networking. We use a pair of in-vacuum aspheric lenses, each with a numerical aperture of 0.8, to couple 10(1)$\%$ of the 493~nm photons emitted from a $^{138}$Ba$^+$ ion into single-mode fibers. We also demonstrate that proximal effects of the lenses on the ion position and motion can be mitigated. 
\end{abstract}

\maketitle

\section{Introduction}\label{sec:Intro}
Trapped ions are a leading platform for quantum computing and networking, with the longest coherence time of any qubit; \cite{wang2021} the highest fidelities for state preparation, measurement, \cite{an2022} and single-qubit gates; \cite{harty2014} and some of the highest fidelity two-qubit gates.\cite{gaebler2016, ballance2016, Srinivas2021, Clark2021}. However, these and other demonstrations of trapped ion quantum computers have all involved small systems with at most tens of qubits. Prominent applications such as Shor's algorithm will likely require at least thousands, if not millions, of qubits.\cite{gidney2021, alexeev2021} Simply adding more ions to a single trapping potential faces several limitations, including spectral crowding of the motional modes\cite{monroe2013} and increased motional heating.\cite{cetina2022}
One option for avoiding these limitations is known as the quantum charge-coupled device (QCCD), which involves multiple trapping zones on a single chip and the ability to shuttle ions between such zones. \cite{kielpinski2002, pino2021} However, the shuttling and cooling overhead may scale poorly with the number of ions in the system.\cite{Moses2023} 

A higher-level modular architecture that can be scaled to larger numbers of ions with full connectivity consists of multiple separate ion traps linked via photonic qubits and reconfigurable optical switches.\cite{monroe2013,Monroe2014} Typically, this involves two distinct types of trapped ion qubits in each module: a communication qubit and a memory qubit.\cite{inlek2017, Feng2023} The communication qubit is used for generating remote entanglement between traps, while the memory qubit is used for local operations and storing quantum information. Because photonic links are lossy, the remote entanglement generation between two modules in this architecture is a heralded process, requiring many attempts to generate entanglement between ions in separate traps. To avoid decoherence of the memory qubit during these attempts, spectrally isolated qubits are envisioned for the communication and memory qubits. Here, we use $^{138}$Ba$^+$ for the communication qubit and $^{171}$Yb$^+$ for the memory qubit.

Ideally, the rate of local operations and remote entanglement should be comparable to facilitate computations utilizing the full resources of the modular architecture. However, recent work has demonstrated a remote entanglement rate of 182~s$^{-1}$, corresponding to a success probability of $2.18\times10^{-4}$. \cite{stephenson2020} This rate is significantly slower than the typical rate for local entanglement between nearby trapped ion qubits.\cite{Wright2019,cetina2022,Moses2023}

One of the primary sources of photon loss and a major limitation to the rate of remote entanglement generation is the finite fraction of light collection. An option for increasing the light collection as well as improving the coupling to the single-mode (SM) fibers necessary for achieving high-fidelity entanglement is the use of cavities placed around the trapped ions. While a cavity can enhance the rate of spontaneous emission \cite{takahashi2020} and increase the probability of successfully coupling a photon into a SM fiber, \cite{schupp2021} systems with trapped ions in cavities have not improved upon the rate demonstrated in Ref.~\onlinecite{stephenson2020}. Additionally, achieving high rates with a cavity usually involves placing high-reflectivity dielectric mirrors very close to the ion. However, uncontrolled charging of the dielectrics can destabilize the ion trap and result in larger heating rates, \cite{teller2021} degrading local entangling operations required for scaling. Single concave mirrors can also be used either by themselves \cite{maiwald2012, chou2017} or in conjunction with a lens \cite{araneda2020} to collect large portions of the fluorescence from the ion. The mirrors used in these systems can have metallic surfaces, reducing the concerns about heating, but high efficiency fiber coupling of light from a trapped ion has not yet been demonstrated with such a system.

We use an alternative approach to increasing the light collection in a trapped ion system. Our system includes two aspheric lenses, each with a numerical aperture (NA) of 0.8. A similar setup has previously been demonstrated in Ref.~\onlinecite{gerber2009}, but with much lower numerical aperture lenses (NA = 0.4) and therefore lower collection efficiencies (a maximum of 8$\%$ total in free space). Together, our NA 0.8 lenses ideally collect 40\% of the 493~nm photons emitted into free space when a barium ion fluoresces. In practice, we observe an $\sim$9\% transmission loss through the lenses, resulting in a net 36\% collection efficiency. Each in-vacuum lens collimates the ion fluorescence, and a second (low NA) aspheric lens outside of the chamber focuses the light into a single-mode, non-polarization maintaining fiber for 493~nm. Collimating the light reduces the sensitivity to imperfections of and alignment to the vacuum window.

In Sec.~\ref{sec:Imaging}, we present the design and characterization of the aspheric collection lenses, both before (Sec.~\ref{sec:PreInstallationTests}) and after (Sec.~\ref{sec:IonImaging}) integration with the rest of the ion trap system. In Sec.~\ref{sec:IonImaging}, we demonstrate the coupling of light from the lenses into a SM fiber with $\gtrsim30\%$ efficiency. In Sec.~\ref{sec:sub_TrapDesign}, we present a trap design that accommodates both dual species operation and the high optical access required for the aspheric objectives. We characterize the performance of the trap with the lenses in close proximity in Sec.~\ref{sec:TrapCharacterization}. We conclude and summarize our results in Sec.~\ref{sec:Conclusion}.

\section{Barium Imaging System Design and Testing}\label{sec:Imaging}

As described in Sec.~\ref{sec:Intro}, the imaging path for light on each side of the trapped ion consists of two aspheric lenses--one high-NA lens inside the chamber, which collects and collimates light from the ion, and a second low-NA lens outside the chamber for focusing collimated light into an SM fiber. In this section, we describe the design of these lenses and discuss the performance of the collective imaging system.

\subsection{Lens Design}\label{sec:Lenses}
Achieving a numerical aperture of 0.8 requires very small working distances from the lens to the object or very large diameter optical elements. 
We use a plano-convex geometry with the flat side facing the ion (object) and a working distance of 6~mm. To achieve an NA of 0.8, this requires a clear aperture of at least 16~mm; we use a lens diameter of 25~mm to provide space for mounting the lens without restricting light collection.
\begin{table}
    \renewcommand{\arraystretch}{1.5}
    \centering
    \begin{tabular}{|c|c|}
        \hline
        \textbf{Parameter} & \textbf{Value for in-vacuum asphere}  \\ 
        \hline
        $t$ & 11.000 mm \\ \hline
        $R$ & 10.367 mm \\ \hline
        $\kappa$ & $-1.059$ \\ \hline
        $A_4$ & $6.238\times 10^{-2}$ \\ \hline
        $A_6$ & $-4.388\times 10^{-4} $ \\ \hline
        $A_8$ & $-7.678\times 10^{-3}$ \\ \hline
        $A_{10}$ & $-4.428\times 10^{-4}$ \\ \hline
        $A_{12}$ & $-5.500\times10^{-3}$ \\ \hline
        $A_{14}$ & $5.822\times10^{-3}$ \\ \hline
        $A_{16}$ & $-1.518\times10^{-3}$\\ \hline
    \end{tabular}
    \caption{Parameters describing the in-vacuum high NA aspheric lens (variables described in text). See Fig.~\ref{fig:AsphereDiagram} for a diagram of the lens.}
    \label{tab:HighNAAsphereParameters}
\end{table}

\begin{figure}[h!]
    \centering
    \includegraphics[width=3.37in]{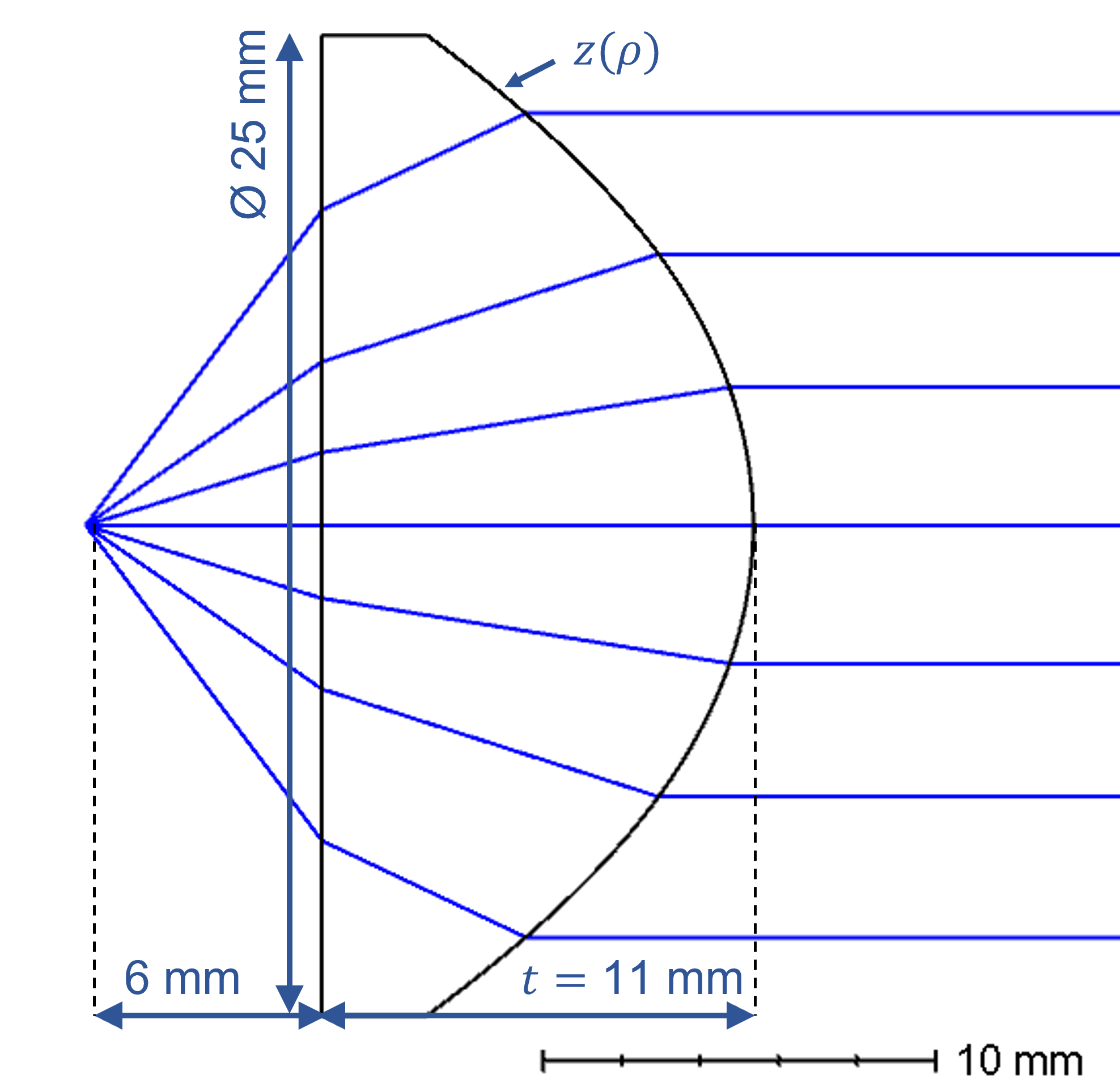}
    \caption{Diagram of the in-vacuum asphere and rays traced from the ion. See Table~\ref{tab:HighNAAsphereParameters} and Eq.~\ref{eq:asphere} for the parameters describing the curved surface.}
    \label{fig:AsphereDiagram}
\end{figure}

For the fiber coupling lens outside of the vacuum chamber, the primary requirement is that it should focus the collimated light from the in-vacuum lens and couple it into an SM fiber. We measure the mode of a typical fiber to have an effective numerical aperture of 0.093. Because of the much lower NA of the fiber coupling lens, it is both easier to manufacture and less sensitive to imperfections. The design, tolerancing, and manufacturing of both lenses were performed by Asphericon, GmbH.

A cylindrically symmetric, aspheric lens surface is typically parameterized by the functional form, \cite{OpticsRef}
\begin{equation}
    \frac{z(\rho)}{R}= \frac{\rho^2}{1+\sqrt{1-(1+\kappa)\rho^2}} + \sum_{\substack{n>2\\n \ even}} A_n \rho^n
    \label{eq:asphere}
\end{equation}
where $z(\rho)$ is the distance along the optical axis, $\rho=r/R$ is the perpendicular distance from the optical axis scaled to the on-axis radius of curvature $R$ of the lens, $\kappa$ characterizes the conic component of the surface, and $A_n$ is the unitless coefficient for the term of order $n$ in the sum. The in-vacuum asphere is made from S-TIH53 glass, which has a refractive index of 1.87 at 486~nm.\cite{Ohara} This relatively high index reduces the required central thickness of the lens to $t=11$~mm and maintains a moderate radius of curvature $R= 10.367$~mm, even with the very large numerical aperture. The conic and polynomial parameters for the aspherical surface facing away from the ion are listed in Table~\ref{tab:HighNAAsphereParameters}. A diagram of the lens with rays traced from the ion is shown in Fig.~\ref{fig:AsphereDiagram}.

The RMS irregularities (RMSi)\cite {OpticsRef} for the two in-vacuum lenses were measured by the manufacturer to be 17 nm and 14 nm on the curved surface and 22 nm and 20 nm on the planar surface, all less than $\lambda/20$ at 493 nm. 

 The curved surface of the lens was coated with an anti-reflection (AR) coating suitable for 493.5~nm. The planar surface was coated with an $\sim$10~nm thick layer of indium tin oxide (ITO), a nearly transparent conductive coating that should shield the ion from static and fluctuating charges on the dielectric surface of the lens.\cite{reens2022} An electrically grounded gold wire was placed near the outside of the lens between the planar surface and the Macor holder and in electrical contact with the ITO layer (described in Sec.~\ref{sec:Optomechanics}). The impact of the lenses on the ion's motion is discussed in Sec.~\ref{sec:TrapCharacterization}. The resulting transmission of the lenses is $91(3)\%$. This value is similar to that measured through a six-element AR-coated objective with a NA of $\sim$0.6 used in earlier trap designs.\cite{crocker2019}

\begin{figure}[t]
    \centering    
    \includegraphics[width=3.37in]{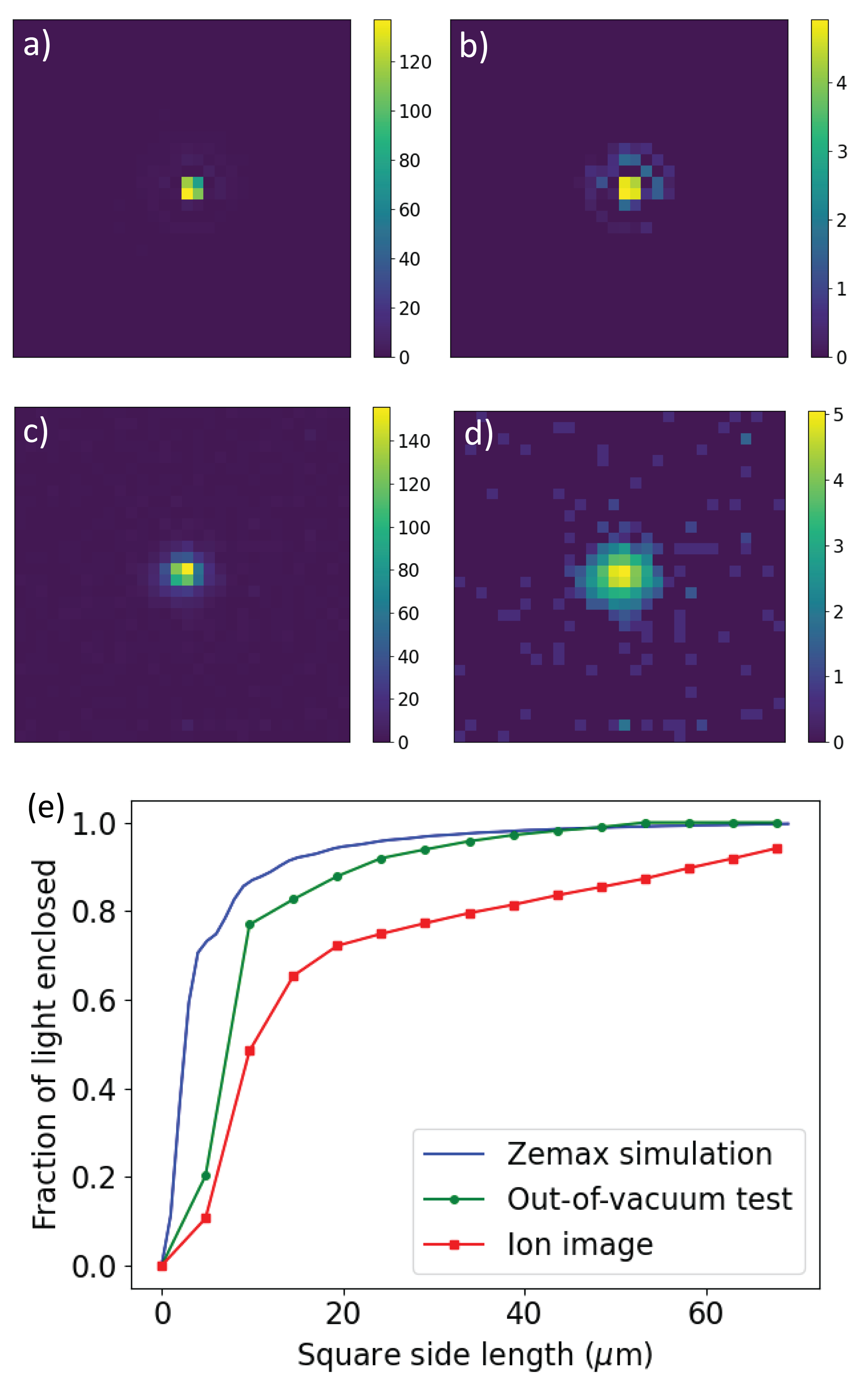}
    \caption{Comparison of ideal and experimentally demonstrated spot sizes. (a) Image of a 200~nm aluminum coated fiber tip. The image was obtained with a CMOS camera with a pixel size of 2.2$\times$2.2~$\mu$m$^2$. (b) Image with a logarithmic color scale of a 200~nm aluminum coated fiber tip. (c) Image of a trapped $^{138}$Ba$^+$ ion [same as Fig.~\hyperref[fig:LensMisalignmentTest]{\ref*{fig:LensMisalignmentTest}(a)}]. (d) Image with a logarithmic color scale of a trapped $^{138}$Ba$^+$ ion. (e) Fraction of the light enclosed in a square centered at the centroid of the image vs the side length of the square for a Zemax OpticStudio calculation (blue line), the image of the fiber tip in (a) (green circles), and the image of the trapped ion in (b) (red squares).}
    \label{fig:SpotSizeAnalysis}
\end{figure}

\begin{figure*}[t]
    \centering
    \includegraphics[width=6.69 in]{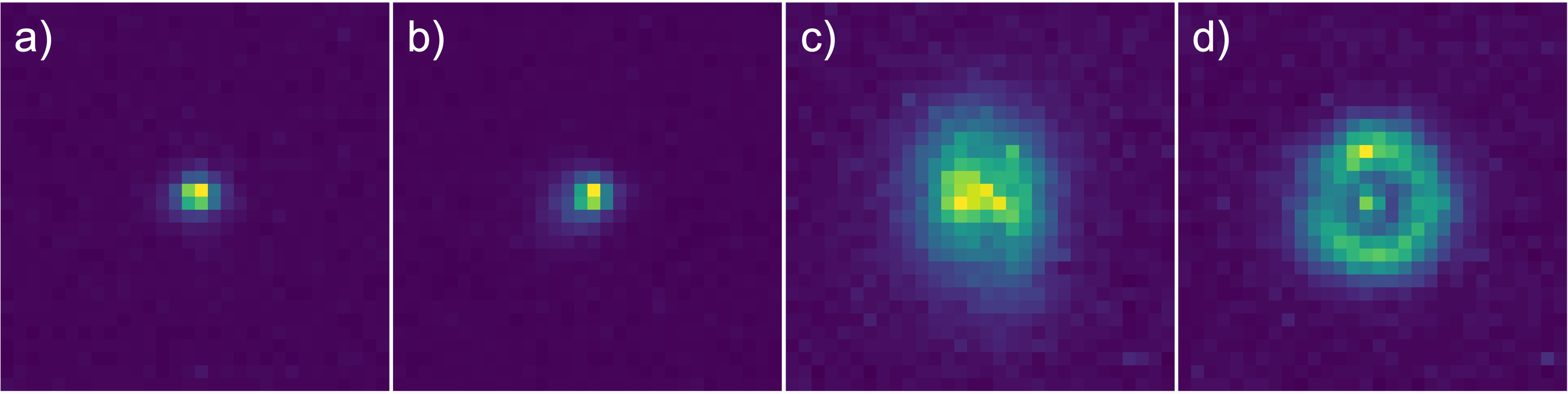}
    \caption{Fluorescence images of a single trapped barium ion through the aspheric lens system. Each image is $66$ $\mu$m on a side. (a) Image of ion with well-aligned lenses. (b) Image of ion with lens displaced transverse to the optical axis by 5~$\mu$m. (c) Image of ion with lens 2~$\mu$m too close to the ion. (d) Image of ion with lens 2~$\mu$m too far from the ion.}
    \label{fig:LensMisalignmentTest}
\end{figure*}

The fiber coupling lens is manufactured from N-BK7 and has a thickness of 10 mm. Both surfaces are curved, although only the surface closer to the ion is not spherical, having a radius of curvature of 68.592 mm and a conic coefficient of -1 (parabolic) with no additional polynomial terms. The second surface is spherical with a radius of curvature of -291.0 mm. 
Because of the lower NA, this lens is easier to manufacture, and thus, the surface irregularities can be limited to an RMSi of 10~nm. Both surfaces of this lens have an AR coating for 493.5~nm.

Zemax simulations of the two-lens system predict a diffraction-limited image and a theoretical maximum fiber coupling of 70$\%$. This maximum value is comparable to that for the lens with NA 0.6 used in Ref.~\onlinecite{crocker2019}. These calculations assume an input apodization of cosine cubed, which corresponds to the intensity distribution from a point source. \cite{laughlin2002}

\subsection{Pre-Installation Lens Testing}\label{sec:PreInstallationTests}

The first tests performed on the lenses were done outside of the vacuum chamber to enable characterization of the performance without possible deleterious effects from the vacuum window. We used an aluminum-coated tapered fiber tip with a diameter of 200~nm as an artificial point source. \cite{robens2017} The high NA asphere was carefully aligned to the fiber tip to minimize aberrations and the fiber coupling asphere was added to reimage the fiber tip onto a CMOS camera with a pixel size of 2.2$\times$2.2~$\mu$m$^2$.

The results of this test are shown in Fig.~\ref{fig:SpotSizeAnalysis}. Fig.~\hyperref[fig:SpotSizeAnalysis]{\ref*{fig:SpotSizeAnalysis}(a)} shows the image of the point source with background counts subtracted. The light is mostly confined to the central four pixels, but a faint ring can be seen outside as well. This ring is shown more clearly in Fig.~\hyperref[fig:SpotSizeAnalysis]{\ref*{fig:SpotSizeAnalysis}(b)}, which shows the same image with a logarithmic color scale. Our analysis for this test consists of calculating the fraction of photon counts within an $N\times N$ pixel$^2$ square. Our data are limited by the fairly large size of the pixels compared to the spot size of the image, but we see in Fig.~\hyperref[fig:SpotSizeAnalysis]{\ref*{fig:SpotSizeAnalysis}(e)} that we can obtain a characterization of the lens performance. In this plot, we show the theoretical enclosed fraction of the light from a Zemax OpticStudio simulation as well as the results of the above analysis for both the test image and an image of a barium ion (see Sec.~\ref{sec:IonImaging}). The performance of the lens in this test is slightly worse than the simulated prediction. We attribute this discrepancy to imperfections in the lens and potential discrepancies between the light emitted from the fiber tip and that from a perfect point source. In Sec.~\ref{sec:IonImaging}, we discuss the resulting fiber coupling with the larger spot size of the ion image.

\subsection{Optomechanical Design}\label{sec:Optomechanics}

The high numerical aperture and aspheric design of the in-vacuum lenses result in a small field of view and depth of focus. As a result, we require alignment capabilities of $\lesssim 100$~nm. As a qualitative demonstration of the sensitivity, we misaligned the lenses by fixed amounts and captured images of single barium ions as shown in Fig.~\ref{fig:LensMisalignmentTest}. Displacements of only a few micrometers, especially along the optical axis, result in large aberrations, which would significantly limit our ability to couple light into a single-mode fiber.

\begin{figure}[h]
    \centering
    \includegraphics[width=3.37in]{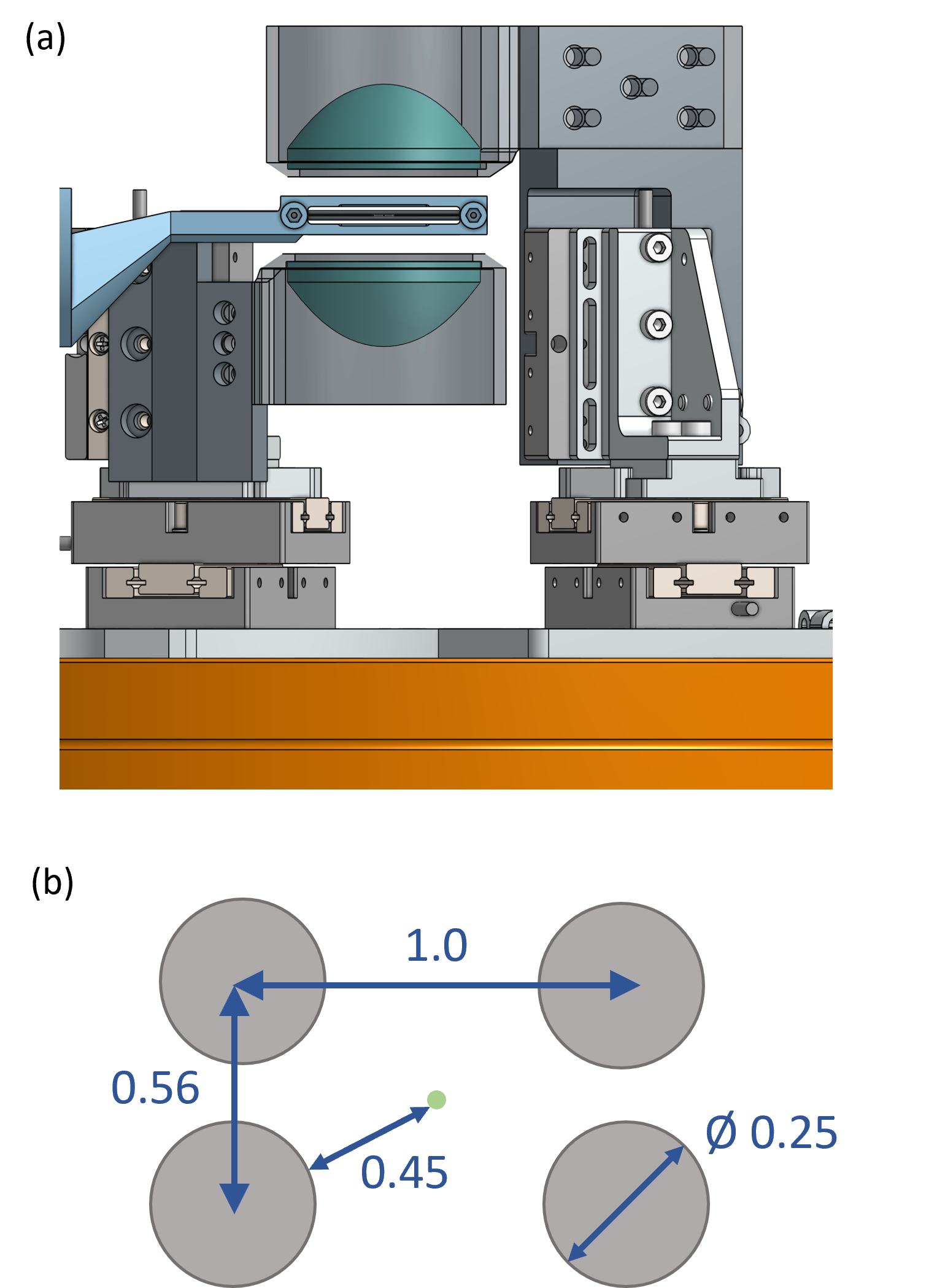}
    \caption{(a) CAD model of the four rod ion trap with aspheres mounted to piezo stages above and below. The ion trap holder is shown in blue, with the rods in the center. The aspheric lenses are teal, and the base of the vacuum chamber is orange. Reproduced with permission from  A. L. Carter, \textit{Design and Construction of a Three-Node Quantum Network}, Ph.D. thesis, University of Maryland, College Park (2021), Copyright 2021 A. L. Carter. (b) Cross section of the four rod ion trap (dimensions in mm). The green point in the center represents the ion. The needles are not shown.}
    \label{fig:AspheresCAD}
\end{figure}

The alignment of the lenses is complicated by the requirement to have them inside an ultrahigh vacuum (UHV) environment. We use piezo stages (Smaract GmbH, SLC-1730-O20-W-S-UHV-NM) that are specified to be compatible with vacuum pressures of $\leq 10^{-11}$~Torr at room temperature and provide step sizes of 1~nm with a closed loop repeatability of 30~nm. These piezos cannot be baked to higher temperatures than 130~$^\circ$C. Nevertheless, with these stages, we are able to achieve a vacuum pressure of $\sim2\times10^{-11}$~Torr, mainly limited by hydrogen outgassing from the stainless steel chamber. 

The layout of the piezo stages with the aspheres and the ion trap is shown in the computer-aided design (CAD) model in Fig.~\hyperref[fig:AspheresCAD]{\ref*{fig:AspheresCAD}(a)}. The ion trap is in the middle of Fig.~\hyperref[fig:AspheresCAD]{\ref*{fig:AspheresCAD}(a)} with the aspheres above and below 6 mm from the ion. To mitigate thermal stress during the process of baking the vacuum chamber, the lens holders (transparent dark gray) and adapter plates (opaque dark gray) are manufactured from Macor, which has a similar coefficient of thermal expansion to the glass of the lenses (S-TIH53).\cite{Ohara, Corning} The lenses are held in place with Macor retaining rings. We also place pieces of indium wire between the retaining rings and the lenses to allow any stresses caused by thermal cycling to deform the indium rather than the glass. The axially symmetric mounting minimizes aberrations from asymmetric stresses on the lens that cannot be reduced by adjusting the lens alignment.\cite{liebetraut2013}

The fiber coupling lenses are also mounted directly to the vacuum chamber to maximize robustness to thermal drifts and other slow changes in the lab that could result in misalignment. This mounting configuration results in stable fiber coupling for days instead of the hours achieved with previous systems with lenses mounted to the optical table. The fiber coupling lenses require only two angular degrees of freedom since the large size of the beam and small divergence angle at the fiber coupling asphere makes it insensitive to misalignments of a few mm in both translational directions perpendicular to the optical axis. Additionally, the Rayleigh range of a Gaussian beam corresponding to the predicted size of the collimated beam from the ions is hundreds of meters, so the distance of the fiber coupling lens can vary by meters without impacting the fiber coupling efficiency. The fiber requires translation in all three dimensions as well as two angular degrees of freedom. All out-of-vacuum adjustments are accomplished using standard optomechanics. 

Overall, the system is aligned by first minimizing aberrations by adjusting the in-vacuum asphere while observing the image of the ion on a camera. Further aberration corrections are then performed with the fiber coupling asphere, although the contributions from this lens are generally much smaller than those from the in-vacuum lens. For optimizing alignment, the aberrations are analyzed using the techniques discussed in Refs.~\onlinecite{Carter2021,wongcampos2016}. Finally, the fiber position and angle are adjusted to maximize the coupling efficiency. This technique allows us to optimize each part of the light collection system nearly independently. The separation of the translation of the in-vacuum asphere from the tilt of the fiber coupling asphere enables much simpler alignment with this system than with a large out-of-vacuum objective, such as in Ref.~\onlinecite{crocker2019} where the tilt and translation degrees of freedom are not independent.

\subsection{Ion Imaging and Fiber Coupling Results}\label{sec:IonImaging}
A similar test to the one presented in Sec.~\ref{sec:PreInstallationTests} can also be performed with a trapped ion as the light source. We calculate the fraction of light enclosed within a square with a varying side length for the image shown in Fig.~\hyperref[fig:LensMisalignmentTest]{\ref*{fig:LensMisalignmentTest}(a)} and Fig.~\hyperref[fig:SpotSizeAnalysis]{\ref*{fig:SpotSizeAnalysis}(c)}. The results are plotted in Fig.~\hyperref[fig:SpotSizeAnalysis]{\ref*{fig:SpotSizeAnalysis}(e)}. For the first two points, the fraction of the light enclosed is substantially smaller than both the simulation and the out-of-vacuum test. We attribute this discrepancy to curvature in the vacuum window, which introduces additional aberrations. 

Ultimately, the most significant test of the lens performance is our fiber coupling efficiency of 30(3)$\%$. We calculate this number by measuring the number of photons on a photomultiplier tube (PMT) placed immediately after a 100~$\mu$m aperture located at the focus of the fiber coupling lens and measuring the number of photons on the same PMT immediately after a fiber. (The 100~$\mu$m aperture blocks only light scattered from the laser beams used to address the ion and essentially no light from the ion itself). We conservatively estimate a systematic uncertainty in this comparison of $\sim$10\% of the measured 30\% due to possible small discrepancies in the experimental setup at the two locations. The observed 30(3)\% fiber coupling efficiency is significantly lower than the simulated prediction of 70$\%$ (see Sec.~\ref{sec:Lenses}). An additional source of photon loss is the dependence of the fiber coupling on photon polarization.\cite{stephenson2020} This accounts for roughly half of the discrepancy between the measured and theoretical fiber coupling. We attribute the remaining efficiency decrease to optical phasefront errors from imperfections in the lenses and vacuum windows and the resulting aberrations. 

To summarize, our overall light collection efficiency for each lens is determined by a product of the $20\%$ acceptance solid angle of the NA 0.8 lens and the above $30(3)\%$ fiber coupling efficiency.  
The overall efficiency is further reduced from clipping on the trap rods [$97(1)\%$ transmission] and loss through the in-vacuum asphere [$91(3)\%$ transmission].
Overall, we couple $5.3(5)\%$ of the 493~nm photons emitted from a single $^{138}$Ba$^+$ ion through a single mode fiber.
This is an improvement over previously reported efficiencies in comparable systems.\cite{stephenson2020}

\section{Ion Trap Design and Performance}\label{sec:TrapDesignandPerformance}
\subsection{Ion Trap Design}\label{sec:sub_TrapDesign}

Achieving high numerical aperture imaging from multiple directions is difficult in many standard ion trap types and impossible even for surface traps designed for high optical access.\cite{maunz2016} The design requirements for this trap are especially stringent since we also plan to perform state readout of $^{171}$Yb$^+$ ions. This operation requires significant light collection to achieve high fidelities.

We use a four rod trap due to its simplicity. To accommodate the larger optical access required for an NA 0.8 lens compared with that required for the NA 0.6 lenses previously used in our system,\cite{crocker2019} we use rods and needles with a 0.25~mm diameter, but maintain a distance from the ion to the rods of $\sim0.45$~mm using a rectangular aspect ratio. In the direction for barium light collection [top and bottom in Figs.~\hyperref[fig:AspheresCAD]{\ref*{fig:AspheresCAD}(a)} and \hyperref[fig:AspheresCAD]{(b)}], the center-to-center distance of the rods is 1~mm, and the distance on the sides of the trap is designed to be 0.56~mm (see Fig.~\hyperref[fig:AspheresCAD]{\ref*{fig:AspheresCAD}(b)}). We use needles to provide axial confinement. The design spacing for the needles was 2.6~mm but by moving an asphere and scattering 493~nm light off the needles, we measure the actual distance between them to be 3.3~mm. With the trap described here, we obtain optical access with a numerical aperture of $0.26$ from a direction orthogonal to the aspheres' optical axis for collection of the 369~nm photons emitted from $^{171}$Yb$^+$. In the directions for collection of 493~nm photons, we calculate that these trap dimensions result in blocking of 3$\%$ of the photons that would otherwise be collected with an 0.8 NA lens. For comparison, the dimensions of the previous four rod traps used in our lab result in a $\sim$20$\%$ loss of photons with an NA~$= 0.6$ objective lens.\cite{sosnova2020} 

\subsection{Trap Characterization}\label{sec:TrapCharacterization}
We use parametric excitation\cite{Zhao2002,Ibaraki2011} to measure the secular frequencies of a single barium ion in the trap. Before compensating micromotion, we observe a change in fluorescence at 330~kHz, 705~kHz, and 888~kHz with 600~V DC applied to the needles, an $\sim$1~kV peak-to-peak RF signal applied to the trap rods, and a DC quadrupole of 1.01~V applied across the trap rods.\cite{wineland1998} When multiple ions are trapped with these voltage settings, the ions align along the trap axis, and thus, the radial confinement is larger than the axial. We can then attribute the lowest frequency of 330~kHz to the axial secular frequency, which is the relevant value for our local interspecies entangling gates.

We investigate the impact of the lenses on the ions using two methods. First, we compensate excess micromotion by suppressing the micromotion sideband of the $\mid 6^2S_{1/2}, m_J=-\frac{1}{2}\rangle$ to $\mid 5^2D_{5/2}, m_J=-\frac{1}{2}\rangle$ transition of a $^{138}$Ba$^{+}$ ion. \cite{blinov2010,yum17} These sidebands are induced by frequency modulation in the ion's reference frame with index 
\begin{equation}
    \frac{\beta}{2}\approx\frac{J_1(\beta)}{J_0(\beta)}=\frac{\Omega_{\pm1}}{\Omega_0} 
\end{equation} 
as micromotion drives the ion along the beam.\cite{Keller2015} We can estimate the modulation index by measuring the carrier $\Omega_0$ and first sideband $\Omega_{\pm1}$ Rabi frequencies. For a carrier $\pi$ time of $\sim$5.5~$\mu$s, we interrogate the micromotion sideband for 500~$\mu$s to look for excess micromotion.

\begin{figure}
    \centering
    \includegraphics[width=3.37in]{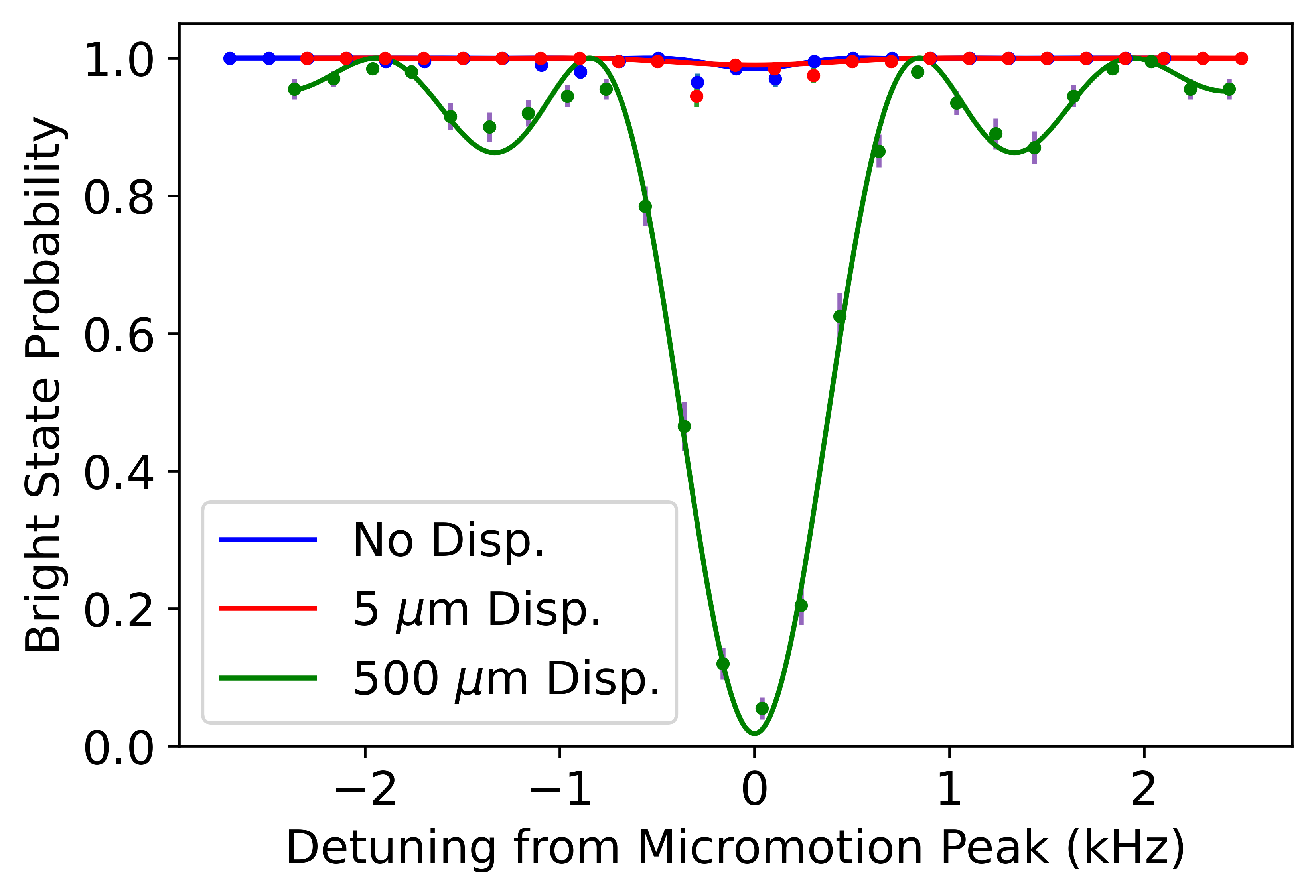}
    \caption{Measured micromotion sideband of the $\mid 6^2S_{1/2}, m_J=-\frac{1}{2}\rangle$ to $\mid 5^2D_{5/2}, m_J=-\frac{1}{2}\rangle$ transition in $^{138}$Ba$^{+}$ at a wavelength of 1762 nm with one in-vacuum asphere in different positions. The interrogation time here is $500~\mu s$ and the carrier $\pi$ time is $\sim5.5~\mu$s. Displacements are measured from the nominal asphere position of about 6 mm from the ion.}
    \label{fig:Micromotion}
\end{figure}

The 1762~nm light to drive this transition is produced and amplified by a thulium-doped DFB fiber laser and fiber amplifier (NKT Photonics Basik and Boostik T20 linecards). We move one in-vacuum lens away from the trap while using the other to continue imaging the ion. As shown in Fig.~\ref{fig:Micromotion}, displacements on the order of the depth of focus of the lens ($<$1~$\mu$m) have a negligible effect ($\beta\ll0.01$), while a much larger displacement of 500~$\mu$m induces small but measurable excess micromotion ($\beta\approx0.022$). Such large adjustments are not necessary after initial alignment of the imaging system.
The smaller displacements are made infrequently, and we can compensate micromotion well at any lens position, so this behavior does not pose a problem for any of our planned experiments.

\begin{figure}
    \centering
    \includegraphics[width=3.37in]{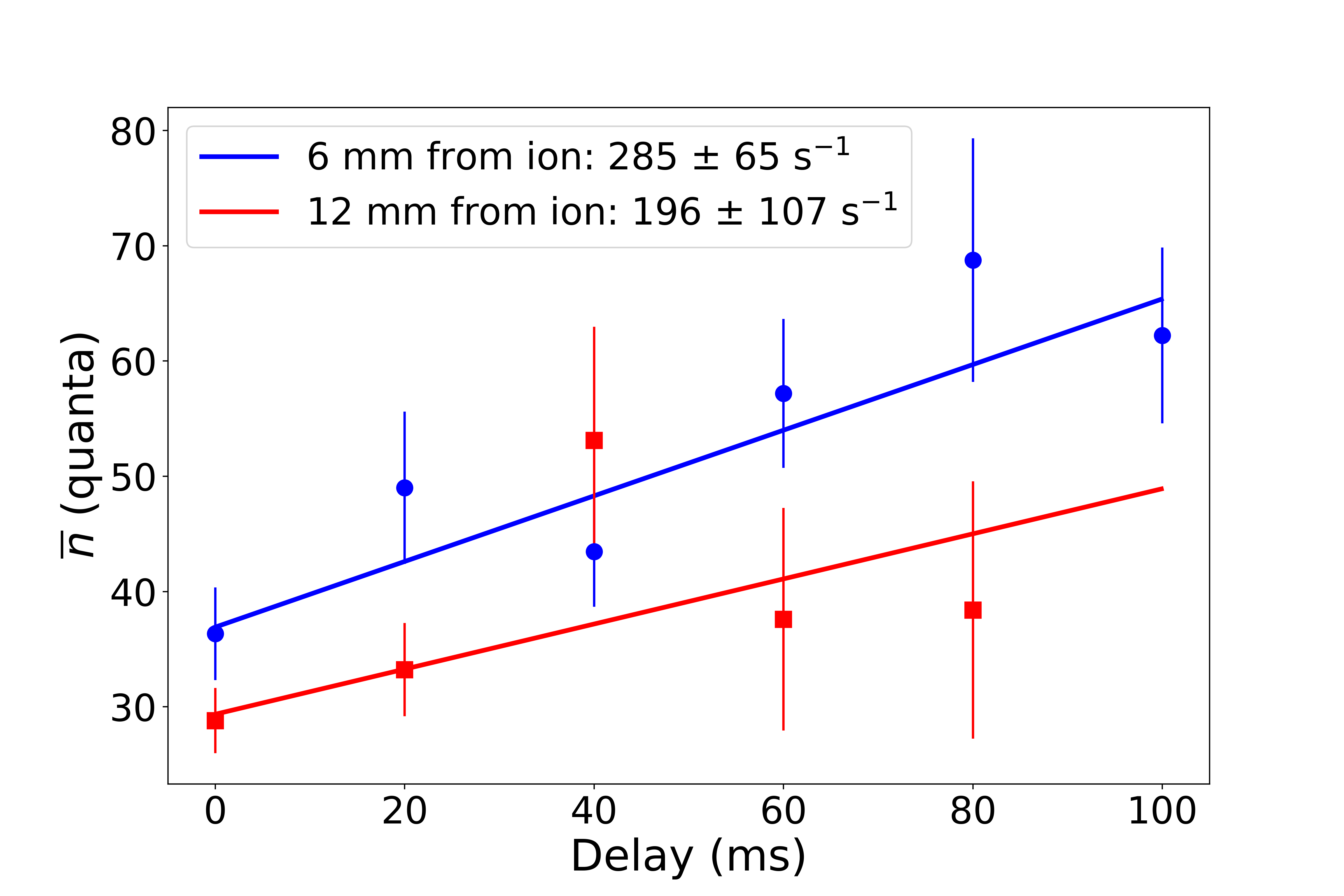}
    \caption{Measured heating rate of a single trapped $^{171}$Yb$^{+}$ ion with the in-vacuum aspheres at their nominal positions (blue circles) and with the aspheres 6~mm farther from the ion than their nominal position (red squares). The straight lines are linear fits to the data. Measurements were performed using the axial motion at a trap frequency of 286~kHz.}
    \label{fig:HeatingRates}
\end{figure}

On the other hand, if charges that build up on the aspheres fluctuate in time, excess heating of the ions could occur, which would lower the fidelity of local entangling operations that rely on the motional modes of the ions. We investigate excess heating of the ion by fitting the decay of carrier Rabi oscillations.\cite{teller2021, haffner2008} For the measurement, we use a trapped $^{171}$Yb$^{+}$ ion so that we can study the effect of moving the barium detection system. We start by Doppler cooling and preparing the ion in the $|6^{2}S_{1/2},\, F=0,\, m_{F}=0 \rangle$ ground state. After waiting some delay time, we drive the ion to the $|5^{2}D_{3/2}, \, F=2,\, m_{F}=0 \rangle$ state with a narrowband 435 nm laser aligned to couple to all three motional modes. This trap has a low axial secular frequency of 286~kHz compared to the radials ($\sim$780~kHz), so we expect that the axial heating should dominate by at least a factor of 7.\cite{Brownnutt2015} These secular frequencies are lower than those previously quoted because of the larger mass of the ytterbium ion. We fit the data to obtain the mean axial phonon number $\overline{n}$ for different delay times. The heating rate is first measured with both lenses at the proper distance to image $^{138}$Ba$^{+}$, $\sim$6 mm from the ion. Then we repeat the measurement with both lenses 12 mm from the ion, about as far as we can move them. 

The results are shown in Fig. \ref{fig:HeatingRates}. With the lenses 6 mm from the ion, we measure a heating rate of 285 $\pm$ 65 quanta/s. With the lenses 12 mm from the ion, a lower rate of 196 $\pm$ 107 quanta/s is measured. The difference measured is much smaller than the scaling suggested in Ref.~\onlinecite{teller2021} for heating induced by dielectric materials, leading us to conclude that our heating rate may not be dominated by such effects from the two aspheric lenses and that the presence of the ITO likely helps shield the ions from the dielectric glass surface. Assuming a two-ion entangling gate time of 200 $\mu$s, we expect this heating rate to induce an infidelity on the order of 2.8 $\pm$ 0.6\%,\cite{ballance2016} which is less than state of the art photonic interconnect fidelities.\cite{stephenson2020} This error can be greatly suppressed by driving the out-of-phase mode rather than the center of mass mode \cite{King1998} or by using radial modes,\cite{Zhu2006} which should make high-fidelity experiments with several entangling gates, such as swapping information onto a memory qubit, feasible.

\section{Conclusion}\label{sec:Conclusion}

We have demonstrated the design and operation of an ion trap and optical system that enable a higher efficiency of free space photon collection from an ion than previously achieved. The ion trap simultaneously allows for optical access for lenses with NA 0.8 in two directions while maintaining optical access in the perpendicular direction for a lens with an NA of $\sim 0.26$. This large amount of optical access enables multi-species operations, including simultaneous collection of photons from Ba$^+$ ions for remote entanglement generation and from Yb$^+$ ions for state readout. 

This system required placing lenses close to a trapped ion, but we have shown that any effects that could reduce the fidelity of future operations are small. Specifically, we show that the contribution of the lenses to micromotion can be compensated by adjusting the trap DC voltages and that if there is an increase in heating of the ions due to the lenses, it is small enough that it will not significantly impact the fidelity of local entangling operations or can be mitigated with the use of alternative motional modes for local entangling operations.

On each side of a barium ion, we collect 5$\%$ of the 493~nm photons emitted for an overall collection efficiency of 10$\%$ of the photons from decays from the $6P_{1/2}$ state to the $6S_{1/2}$ state. The enhanced light collection efficiency compared to previous experiments could improve various quantum operations, including state detection,\cite{crain2019} quantum teleportation between stationary and photonic qubits,\cite{arenskotter2023} and information storage in a nearby ion used as a memory qubit.\cite{drmota2023} The inclusion of two aspheres also enables light collection in two opposing directions from an ion, which is advantageous for integration into a trapped ion modular quantum computer or quantum network. For a two-node quantum network using two of the systems described here, the multidirectional light collection along with the increased collection efficiency would increase the remote entanglement generation rate by a factor of 3.5 compared with the current highest rate.\cite{stephenson2020} For a three-node network, the light collected from each asphere can be used to generate entanglement with ions in a different ion trap. The geometry of this system also then enables parallel remote entanglement attempts between multiple traps without excess losses from splitting the light. In addition to increasing the rates for quantum networking, this architecture naturally lends itself as a quantum repeater between two other traps.\cite{Santra2019} Overall, the improvements in photon collection we have demonstrated enable significantly increased rates for various applications of remote entanglement and quantum communication.

\begin{acknowledgments}
This work is supported by the NSF STAQ Program (PHY-1818914), the DOE Quantum Systems Accelerator (DE-FOA-0002253), the ARO MURI on Modular Quantum Circuits (W911NF1610349), the AFOSR MURI on Dissipative Quantum Control (FA9550-19-1-0399), and the AFOSR MURI on Interactive Quantum Computation and Communication Protocols (FA9550-18-1-0161). J.O. is supported by the National Science Foundation Graduate Research Fellowship (DGE 2139754).
\end{acknowledgments}

\section*{Author declarations}

\subsection*{Conflict of interest}

C.M. is a co-founder of IonQ, Inc. and has a personal financial interest in the company.

\subsection*{Author contributions}

\textbf{Allison L. Carter:} Conceptualization (equal); Formal analysis (equal); Investigation (lead); Methodology (lead); Project administration (lead); Supervision (supporting); Visualization (lead); Writing -- original draft (lead); Writing -- review \& editing (equal).
\textbf{Jameson O'Reilly:} Conceptualization (supporting); Formal analysis (equal); Funding acquisition (supporting); Investigation (supporting); Methodology (supporting); Project administration (supporting); Software (supporting);  Visualization (supporting); Writing -- original draft (supporting); Writing -- review \& editing (supporting).
\textbf{George Toh:} Data curation (lead); Formal analysis (supporting); Investigation (supporting); Methodology (supporting); Software (lead); Visualization (supporting); Writing -- original draft (supporting); Writing -- review \& editing (supporting).
\textbf{Sagnik Saha:} Data curation (supporting); Investigation (supporting); Software (supporting); Writing -- review \& editing (supporting).
\textbf{Mikhail Shalaev:} Investigation (supporting); Methodology (supporting); Writing -- review \& editing (supporting).
\textbf{Isabella Goetting:} Investigation (supporting); Software (supporting); Writing -- review \& editing (supporting).
\textbf{Christopher Monroe:} Conceptualization (equal); Funding acquisition (lead); Project administration (supporting); Resources (lead); Supervision (lead); Writing -- review \& editing (equal).

\section*{Data availability}

The data that support the findings of this study are available from the corresponding author upon reasonable request.

\bibliography{aipsamp}

\end{document}